# Utilisation de la substitution sensorielle par électro-stimulation linguale pour la prévention des escarres chez les paraplégiques. Etude préliminaire

A. Moreau-Gaudry[1], F. Robineau[2], P.F. Andre[3], A. Prince[3], P. Pauget[3], J. Demongeot[2], Y. Payan[2].

[1] *Centre Hospitalier Universitaire*
*Grenoble, France*
Alexandre.Moreau-Gaudry@imag.fr

[2] *Laboratoire TIMC-IMAG-CNRS*
*Grenoble, France*
Fabien.Robineau@imag.fr,
Jacques.Demongeot@imag.fr,
Yohan.Payan@imag.fr

[3] *Centre Médico Universitaire Daniel Douady*
*St hilaire du Touvet, France*
PaugetP@club-internet.fr
AnneNicolePrince@aol.com
ergo-cmudd@fsef.net

**Résumé :** *Cet article présente l'étude préliminaire sur des sujets sains d'un nouveau dispositif de prévention des escarres chez les blessés médullaires paraplégiques. Ce dispositif repose sur le principe de la substitution sensorielle, principe qui consiste à suppléer les stimuli d'une modalité sensorielle déficiente (sensibilité fessière) par ceux d'une autre modalité fonctionnelle (sensibilité linguale). Il est constitué de deux parties : une matrice de 36 électrodes (6*6) disposées sur la langue et connectées à un ordinateur portable; un ensemble de capteurs de pression, connectés au même ordinateur et disposés dans un coussin sur lequel le sujet va s'asseoir. L'expérimentation a consisté à activer, par stimulation électrique les récepteurs tactiles de la langue selon une « image directionnelle » et à enregistrer les variations de pression au niveau du coussin, le patient devant modifier sa posture selon l'information de direction décodée. Les résultats obtenus montrent, chez 10 sujets sains, dans 92% des cas, une réponse en pression adaptée à l'image directionnelle stimulée. Ces résultats permettent d'envisager l'utilisation de ce dispositif à la prévention des escarres de pression chez les sujets paraplégiques, par identification des zones de souffrance tissulaire à partir d'une analyse automatique par l'ordinateur de la carte de pression, et stimulation électrique linguale d'une direction de mobilisation posturale à adopter par le paraplégique pour s'affranchir de cette surpression.*

## 1. Introduction

### 1.1. Escarre & Paraplégie

Une escarre est définie comme une « nécrose tissulaire d'origine ischémique formant une croûte noirâtre plus ou moins épaisse tendant à s'éliminer et surtout à s'infecter » [Garnier, 01]. Privés de leurs capacités sensorielles et motrices, les lésés médullaires paraplégiques, assis dans leur fauteuil, ne possèdent plus le réflexe inconscient de changer de position comme le fait une personne valide assise. Ils développent ainsi des escarres, aux conséquences parfois dramatiques, escarres localisées principalement dans la région ischiatique, par surpression locale prolongée. En effet, une pression de 60 à 80 mmHg, maintenue pendant 2 à 3 heures, peut provoquer des escarres [Kosiak, 76 ; Maklebust, 87].

## 1.2. Solutions actuelles

Le contrôle de la charge tissulaire indispensable à la prévention des escarres a engendré le développement de coussin et de matelas "anti-escarre". Ces matériels permettent, par diminution des pressions à l'interface fesses-coussins, de diminuer les forces qui agissent à l'interface muscle-ischion [Hedrick-Thompson, 92]. L'objectif de ces matelas/coussins est d'obtenir une pression inférieure à 32 mmHg, pression considérée comme suffisamment faible pour ne pas provoquer une plaie de pression. Ces matelas réducteurs de pression, constitués de gel, de mousse, d'eau ou d'air, font partie des stratégies actuelles de prévention des plaies dans les milieux hospitaliers. Toutefois, ces supports ne suffisent pas et les techniques de transfert de poids constituent toujours le meilleur moyen de prévention. Soulèvements, changements de position, modifications des appuis sont à réaliser régulièrement (environ toutes les deux heures) afin d'assurer une bonne répartition pressionnelle et ainsi prévenir la formation de cette lésion [LETTS, 95]. Des dispositifs récemment développés [Xsensor Technology Corporation, 05] détectent, au moyen de capteurs de pression positionnés sous les fessiers, les régions en surpression trop longuement exposées (plus de deux heures) et renvoient cette information au patient et au médecin sous forme d'une carte de couleur ou plus simplement par un moyen externe (signal sonore ou lumineux par exemple). Les zones à risque sont ainsi identifiables. Néanmoins, ce dispositif présente deux inconvénients majeurs : il limite la mobilité du patient et le contraint à être attentif en permanence à l'écran de contrôle (au signal sonore ou au signal lumineux). Un des objectifs de cette recherche est de s'affranchir de ces inconvénients par la mise en oeuvre du principe de la substitution sensorielle.

## 1.3. Principe de la substitution sensorielle

Au cours de leurs recherches, Bach-y-Rita et ses collaborateurs [Bach-y-Rita et al., 69 ; Bach-y-Rita, 03] ont montré que des stimuli caractéristiques d'une modalité sensorielle (la vision) pouvaient être remplacés par des stimuli d'une autre modalité sensorielle (le touché). Le premier développement d'un système de substitution visuo-tactile a été réalisé dans le but de fournir une information visuelle distale aux aveugles [Collins et Bach-y-Rita, 73]. Ce système, le « Tactile Vision Substitution System » (TVSS) était composé d'une matrice de 400 électrodes tactiles (20*20) placées sur différentes parties du corps comme le front ou l'abdomen et permettait la traduction, en "images tactiles", d'images capturées par une caméra. Bach-y-Rita et ses collaborateurs ont montré, grâce à ce système, que les aveugles étaient capables de reconnaître des formes du monde extérieur à travers la modalité tactile [Bach-y-Rita, 72]. Après un temps d'adaptation, le sujet aveugle oublie les stimulations sur la peau et perçoit les objets comme étant devant lui. Après plusieurs semaines d'apprentissage, des potentiels évoqués étaient observés au niveau de l'aire visuelle jamais activée auparavant. Autrement dit, les aveugles, sujets aux expérimentations de substitution sensorielle, commençaient par percevoir l'électro-stimulation comme une activité tactile classique, puis se mettaient véritablement à "voir" à travers la paume, les doigts ou le dos.

## 1.4. Modalité de la substitution sensorielle

Dans le cadre de cette recherche, nous nous sommes orientés vers la langue, comme organe de substitution sensorielle, pour deux raisons. La première est en rapport avec les travaux de Bach-y-Rita sur la substitution sensorielle, au cours desquels il a récemment convergé vers cet organe [Bach-y-Rita et al., 98]. En effet, la langue, organe du corps humain possédant la plus grande densité de récepteurs tactiles, dispose d'une capacité discriminative supérieure à celle du bout des doigts [Sampaio, 01]. La seconde est la conséquence directe des traumatismes médullaires : alors que la plupart des centres nerveux sensitifs et moteurs rachidiens peuvent être lésés chez les traumatisés rachidiens, la langue se trouve souvent être une des rares structures préservées.

### 1.5. Objectif

Notre projet de recherche vise à compenser les déficits sensoriels et cognitifs des paraplégiques par l'utilisation d'un dispositif d'électro-stimulation linguale (ensemble d'électrodes placées en bouche) couplé à des capteurs de pression inclus au sein d'un coussin de pression, dans le but de réduire les escarres touchant les régions ischiatiques. Dans cette étude préliminaire, nous nous sommes intéressés à la faisabilité d'une telle approche chez des sujets sans lésion médullaire.

## 2.   Matériels & Méthodes

### 2.1. Coussin de pression

Nous avons développé, en collaboration avec la société Léas®, un coussin de pression (prototype) qui permet la mesure, en temps réel, des pressions appliquées sur les fessiers du paraplégique lorsque ce dernier est assis dessus. Ce coussin est constitué de deux hémi-coussins droit et gauche (cf. Figure1). Chaque hémi-coussin comporte 6x12 capteurs de pression. Chaque capteur de pression a une surface d'environ 1cm². La répartition des capteurs dans le coussin est non linéaire afin de permettre une meilleure résolution dans les zones ischiatiques où les risques d'escarres sont plus élevés. Ces capteurs fonctionnent grâce à une poudre semi-conductrice répartie uniformément dans une enveloppe de polymère. Cette poudre possède des propriétés élastiques et agit selon le principe de percolation : lorsque son volume change sous l'effet d'une force de pression, sa conductivité augmente et la résistance au sein du capteur baisse. La variation de courant envoyé à travers la poudre est ainsi fonction de la pression exercée. Chaque capteur est relié à un système électronique qui permet la mesure de son potentiel électrique codé sur 4 bits (16 niveaux de pression par capteur).

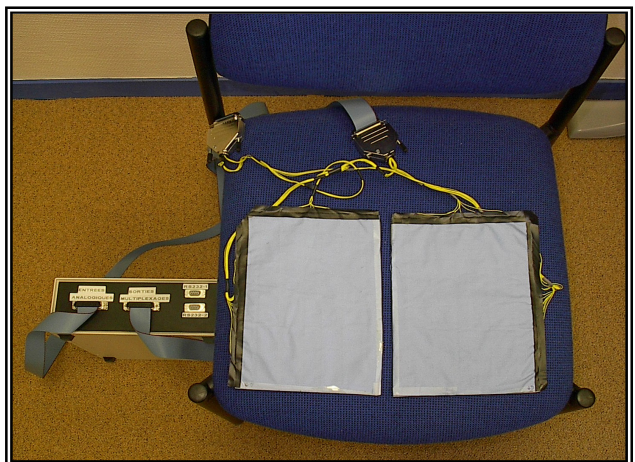

*Figure 1 : Le coussin de pression (prototype).*

## 2.2. Le TDU

Un prototype, nommé « Tongue Display Unit » (TDU) a été développé et validé sur une population de non voyants pour lesquels les informations visuelles captées par caméra vidéo étaient transcodées en stimulations électrotactiles linguales [14]. Dans cette étude, nous utilisons une version simplifiée de ce prototype. Il est constitué d'une matrice de 36 électrodes tactiles (6×6) collées sur une bande plastique et connectées à un système électronique externe (cf. Figure 2). Le sujet peut maintenir les électrodes au contact de sa langue, la bouche fermée. La salive offrant une bonne conductivité, le TDU requiert seulement un voltage de sortie de 5 à 15V et un courant de 0.4 à 4mA pour stimuler les récepteurs linguals. Ainsi, lorsqu'une électrode est activée, le sujet ressent, à ce niveau, un « picotement » à la surface de sa langue. Dans la présente étude, nous avons maintenu constante la fréquence de stimulation (50 Hz), seule l'intensité de la stimulation variait en fonction de la sensibilité du sujet.

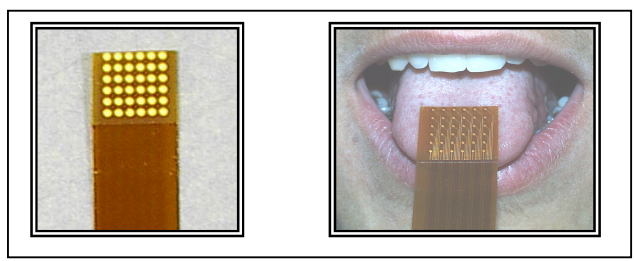

*Figure 2 : Les 36 électrodes du TDU sont placées sur et au contact de la langue.*

## 2.3. La méthodologie

L'objectif est de montrer la capacité d'un sujet, sans lésion médullaire, à percevoir et interpréter l'information fournie par les électrodes du TDU placées sur la langue, puis adopter une attitude posturale adaptée à l'information décodée (à terme, en vue d'une limitation des zones de surpression). Pour ce faire, 10 sujets sains volontaires du laboratoire TIMC (5 de sexe féminin), après signature d'un consentement éclairé, d'âge moyen 26.2 ans (âge minimal : 24 ans, âge maximal : 37 ans) ont accepté de participer à une expérience clinique constituée de trois parties : la première a consisté à *calibrer* la stimulation électrique linguale, du fait de l'anisotropie spatiale individuelle de la sensibilité linguale. Pour chaque individu, cette calibration a été effectuée, à fréquence de stimulation constante (50Hz), en modulant l'intensité de la stimulation linguale en fonction de la perception rapportée par le sujet. La seconde partie était consacrée à l'*apprentissage*. Il a été demandé au sujet d'identifier le groupe d'électrodes activées parmi 4 groupes. Ces groupes étaient constitués des 6 électrodes des bords respectivement antérieur, postérieur, droit et gauche de la matrice d'électrode, dans le référentiel du patient (cf. Figure 3). Enfin, la dernière partie regroupait 10 *épreuves cliniques*, chaque épreuve s'étant déroulée de la manière suivante : la carte de pression de chaque sujet assis sur le coussin de pression, au repos, a été acquise pendant une durée de 3 secondes. Une des 4 électro-stimulations linguales a ensuite été envoyée sur le TDU. Après avoir identifié la région de la matrice activée, le sujet devait mobiliser son buste selon la direction indiquée par l'électro-stimulation linguale (les stimulations antérieure, postérieure, gauche, droite correspondaient respectivement à une demande de déplacement du buste en avant, en arrière, à gauche, à droite). Après 3 secondes, temps accordé au sujet pour mobiliser son buste selon l'information directionnelle reçue sur la langue, la nouvelle carte de pression du sujet, assis, au repos, a été enregistrée. Un score unitaire ou nul a été attribué à chaque épreuve si et seulement si le déplacement des barycentres des cartes de pression, avant et après mouvement du buste, concordait ou non avec l'information linguale électro-stimulée. Ainsi, pour chaque individu, nous avons obtenu un score total côté sur 10. Tous les sujets ont réalisé les mêmes épreuves cliniques.

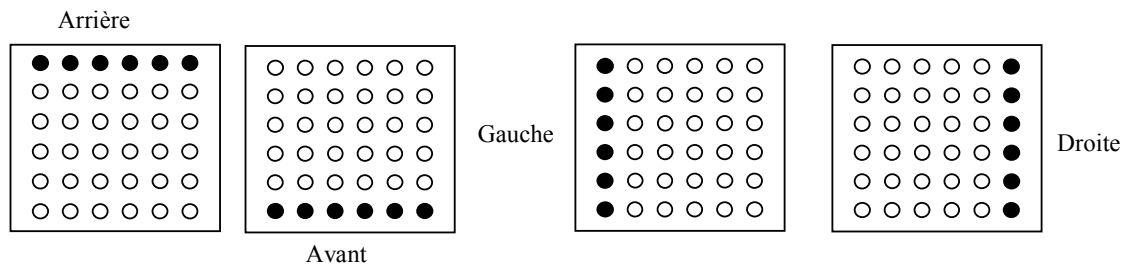

Figure 3 : Schéma des quatre stimulations envoyées sur le TDU pour indiquer la direction de mobilisation du buste à adopter (en arrière, en avant, sur la gauche, sur la droite).

### 3.   Résultats & Discussion

Chaque sujet, ayant accepté de participer à cette épreuve expérimentale, l'a mené à terme : l'étape de calibration individuelle n'a présenté aucune difficulté, l'apprentissage était obtenu au bout de 4 à 8 stimulations, les épreuves cliniques se sont déroulées sans incident. Le score moyen obtenu était de 9.2, avec un écart type de 0.79, un minimum à 8 et un maximum à 10.

Cette étude préliminaire montre, chez des patients sans lésion médullaire, la faisabilité et la fonctionnalité (92% de réussite) de la boucle «stimulation linguale TDU - réaction posturale – évaluation» à partir d'informations pressionnelles fournies par le coussin de pression.

La prochaine étape consistera à l'évaluation de cette boucle par 20 sujets paraplégiques en fauteuil roulant du Centre Medico Universitaire Daniel Douady (CMUDD) de Saint Hilaire du Touvet (38). L'objectif recherché sera l'uniformisation de la répartition des pressions ischiatiques des sujets au cours du temps, le TDU leur indiquant, à partir d'informations pressionnelles, les modifications de posture à entreprendre pour éviter l'apparition de surpression ou pour corriger des zones de surpression.

A terme, nous pensons que les sujets utilisant le système TDU parviendront à adopter, par réflexe inconscient, des mouvements de compensation de la même façon qu'une personne valide. Autrement dit, une prise en compte de type « bas niveau » des mouvements à effectuer en fonction de l'information de posture transmise par le TDU en vue d'une normalisation des pressions ischiatiques.

Plus généralement, nous pensons, comme Bach-y-Rita avec la substitution sensorielle chez les aveugles, que les lésés médullaires pourront "ressentir" les pressions exercées sur leurs ischions à partir de cartes de pression "affichées" sur leur langue, cartes de pression brutes ou prétraitées à partir desquelles ils adapteront leur posture en conséquence.

Afin d'améliorer l'ergonomie et l'utilisation du TDU par une personne en fauteuil roulant, nous développons actuellement une version sans-fil *(communication HF)* et miniaturisée de notre système TDU. Ce prototype sera inséré dans une prothèse palatale (cf. Figure 4 – maquette du dispositif). Ce travail est réalisé en collaboration avec l'entreprise Coronis-Systems (https://www.coronis-systems.com).

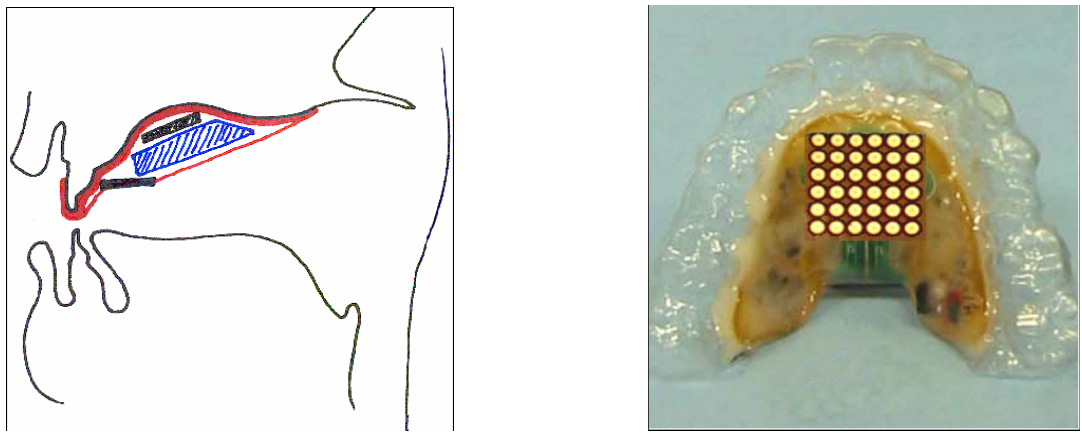

*Figure 4 : schéma de la prothèse palatine avec maquette du prototype en cours de développement.*

Enfin, afin de disposer de données pressionnelles absolues et non plus relatives, un travail de calibration est actuellement en cours.

**4.  Conclusion**

Chez 10 sujets, sans lésion médullaire, la réponse en pression enregistrée au niveau des fessiers suite au déplacement du buste est adaptée à l'information directionnelle préalablement électro-stimulée. Ces résultats permettent d'envisager l'utilisation de ce dispositif à la prévention des escarres de pression chez les sujets paraplégiques, par identification des zones de souffrance tissulaire à partir d'une analyse automatique par l'ordinateur de la carte de pression, et électro-stimulation linguale d'une direction de mobilisation posturale à adopter par le patient paraplégique pour corriger cette surpression.